\documentclass[11pt, conference]{IEEEtran}
\usepackage{epsfig,graphicx,psfrag}
\usepackage{amsfonts,amsmath,amssymb}

\newtheorem{theorem}{Theorem}[section]

\begin{document}

\title{State Information in Bayesian Games}

\author{
\authorblockN{Paul Cuff}
\authorblockA{Department of Electrical Engineering \\
Princeton University\\
E-mail: cuff@princeton.edu }
}

\maketitle

\begin{abstract}
Two-player zero-sum repeated games are well understood.  Computing the value of such a game is straightforward. Additionally, if the payoffs are dependent on a random state of the game known to one, both, or neither of the players, the resulting value of the game has been analyzed under the framework of Bayesian games. This investigation considers the optimal performance in a game when a helper is transmitting state information to one of the players.

Encoding information for an adversarial setting (game) requires a different result than rate-distortion theory provides. Game theory has accentuated the importance of randomization (mixed strategy), which does not find a significant role in most communication modems and source coding codecs. Higher rates of communication, used in the right way, allow the message to include the necessary random component useful in games.
\end{abstract}

\section{Introduction}
\label{section introduction}

\begin{figure}
\psfrag{a}[][][0.9]{$3$}
\psfrag{b}[][][0.9]{$0$}
\psfrag{c}[][][0.9]{$0$}
\psfrag{d}[][][0.9]{$1$}
\psfrag{e}[][][0.9]{0}
\psfrag{f}[][][0.9]{1}
\psfrag{g}[][][0.9]{0}
\psfrag{h}[][][0.9]{1}
\psfrag{i}[][][1]{Player B}
\psfrag{j}[][][1]{Player A}
\psfrag{k}[][][0.9]{$p_A(a)$}
\centering
\includegraphics[width=.3\textwidth]{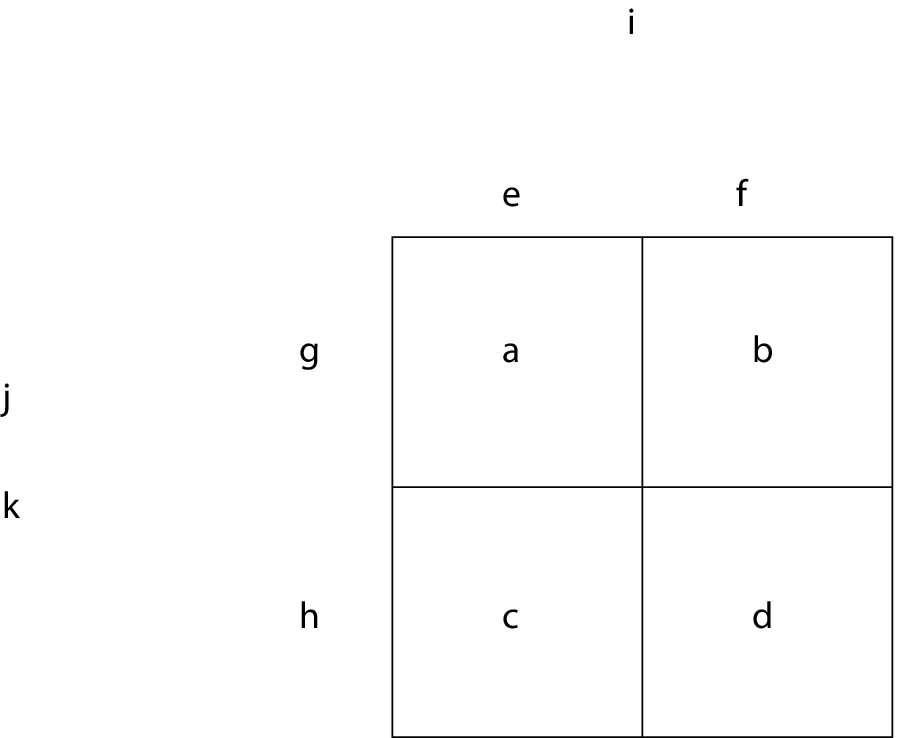}
\caption{{\em Example of two-player zero-sum game.}  The entries in the matrix represent the payoff that Player A receives, which is negative of the payoff for Player B.  To maximize the minimum expected payoff, Player A chooses action 0 with probability 1/4 and action 1 with probability 3/4.  This results in an expected payoff of 3/4 for Player A.}
\label{figure simple game}
\end{figure}

Two-player zero-sum games play a fundamental role in game theory because their analysis is straightforward.  The min-max theorem of von Neumann \cite{vonNeumann1928} establishes a unique value of the game.  Figure \ref{figure simple game} shows the payoff matrix for a basic game.  Since neither of the pure strategies for Player A dominates the other, a mixed strategy is optimal.  Player A will choose action 0 with probability 1/4 and action 1 with probability 3/4.  By doing this, the expected payoff for Player A is 3/4 no matter which strategy Player B chooses.

In a Bayesian game, the payoff matrix, which we represent by $\Pi(a,b)$ over the domain of pure strategies $a$ and $b$, is a random quantity.  We can model this as a game with a random state $S$ that determines the payoffs, referred to in the literature as the ``type.''  The value of such a game depends on the information known to the players.  If the neither player knows $S$ (aside from its distribution), then the value of the game can be derived from the expected value of the payoff matrix.  Additionally, if both players know $S$ then the value can be derived from the payoff matrix associated with every instance of $S$ and averaged.  More interesting cases occur when only one player knows $S$ or when both players have incomplete information about $S$.

\begin{figure}
\psfrag{l1}[][][0.8]{Distribution of state $S$}
\psfrag{l2}[][][0.8]{Quantization into bins}
\centering
\includegraphics[width=.4\textwidth]{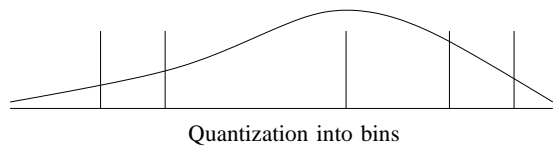}
\caption{{\em Example of information structure.}  This model for discussing partial state (``type'') information in a Bayesian game uses a partition over the domain of the state to represent the ``information structure.''  A player of the game observes only a discrete function of the state indicating the cell that the state belongs to.  Each player of the game may have a different information structure.}
\label{figure information structure}
\end{figure}

Consider a situation where different functions of the state $S$ are known to both of the players of the game.  These functions can be represented by a partition over the support of $S$ known as the information structure, as illustrated in Figure \ref{figure information structure}.  It has been established that games of this form can be solved by expanding the space of pure strategies to include strategies that depend on the available information.  The min-max theorem still holds.  Samples of inquiries into the value of information can be found in the following publications:  \cite{Gossner2006}, \cite{Gilboa1991}, \cite{Shmaya2006}, \cite{Lehrer2006}.  While information can only increase a players optimal score in a game, not all information structures are equal, even if they contain the same quantity of information.  For a given resolution, what is the optimal information structure?

In this work we consider a repeated game setting where the state of the game $S$ is i.i.d. and known non-causally to a helper who is assisting one of the players of the game.  The helper can communicate with a rate limit of $R$ bits per iteration of the game.  The opposing player may or may not know $S$ but is certainly aware of the conspiracy against him.  Even the protocol for communication is known (or learned) by the opposing player.  We establish information theoretic lower-bounds to the optimal performance.  We show through examples that the intuition provided by rate-distortion theory can be misleading in this setting.

\section{Example - Erasure Game}
\label{section example}

\begin{figure}
\psfrag{a}[][][0.8]{3}
\psfrag{b}[][][0.8]{0}
\psfrag{c}[][][0.8]{0}
\psfrag{d}[][][0.8]{1}
\psfrag{e}[][][0.8]{$-\infty$}
\psfrag{f}[][][0.8]{$-\infty$}
\psfrag{g}[][][0.8]{0}
\psfrag{h}[][][0.8]{1}
\psfrag{i}[][][0.8]{0}
\psfrag{j}[][][0.8]{e}
\psfrag{k}[][][0.8]{1}
\psfrag{l}[][][0.8]{Player B}
\psfrag{m}[][][0.8]{Player A}
\psfrag{n}[][][0.8]{}
\psfrag{o}[][][0.8]{$S=0$}
\psfrag{p}[][][0.8]{$S=1$}
\centering
\includegraphics[width=.4\textwidth]{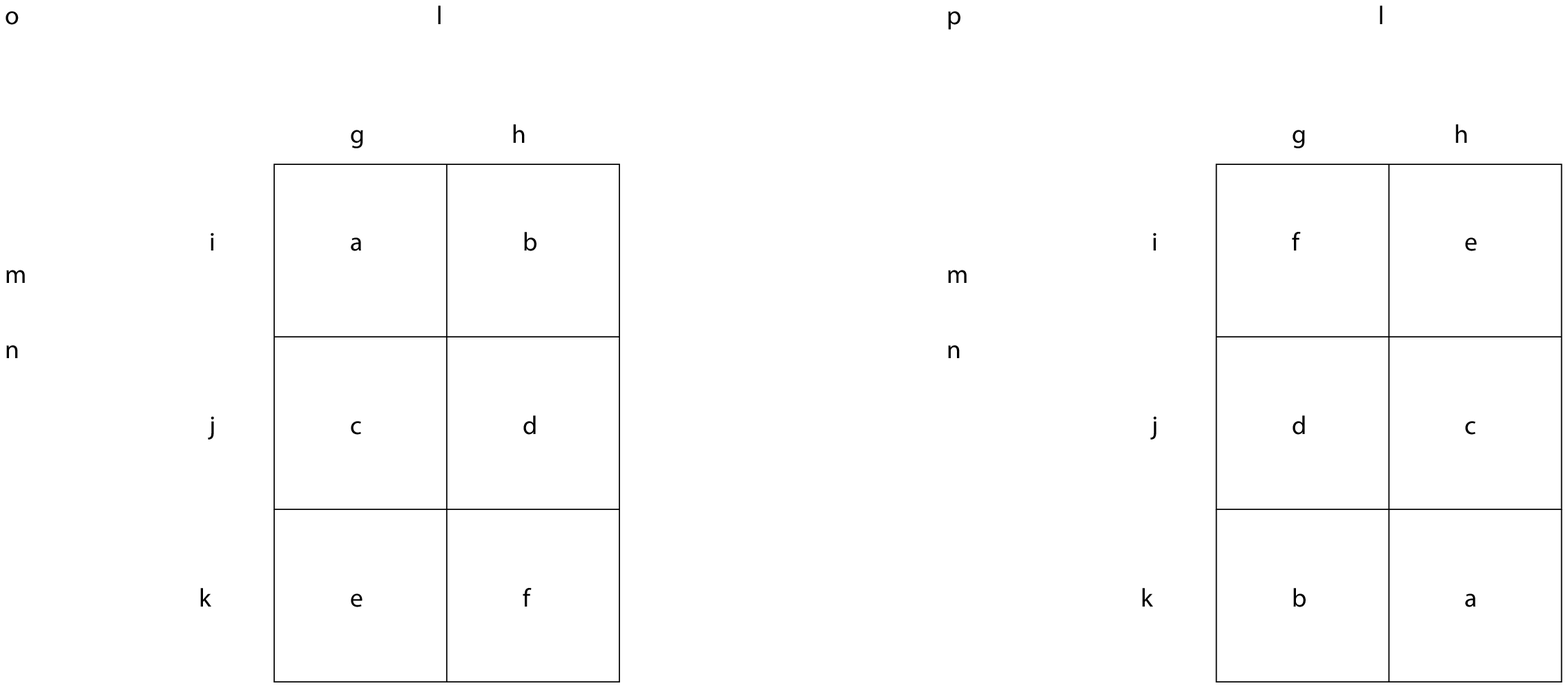}
\caption{{\em Erasure game.}  This game has two states that are equally likely.  Player A must play $A=S$ or $A=e$ with probability 1.  Therefore, if Player A does not know the state, $A=e$ is the only choice.}
\label{figure erasure game}
\end{figure}

We illustrate in Figure \ref{figure erasure game} a two-player game with a binary, equally distributed state.  Player A has three pure strategies, $0$, $e$, and $1$.  Player B only has two strategies, $0$, and $1$.  The matrix values represent the payoff Player A receives for any given state $S$ and pair of actions $A$ and $B$ (pure strategies).  As this is a zero-sum game, Player B receives negative of the payoff of Player A.  We concern ourselves with average payoff, averaged with respect to probabilities if mixed strategies are involved.

Notice that Player A will at all costs avoid choosing a pure strategy of $A=1$ when the state is $S=0$, or vice versa, hence the label ``erasure game.''\footnote{The results in this paper actually require the payoffs to be finite.  Here we use $-\infty$ casually.  The same consequences would result from a finite but greatly negative payoff in place of $-\infty$.}  So if Player A is ignorant of the state $S$, he has no option but to choose $e$ with probability one.  We can therefore deduce the value of the game for two of the four information structures in Table \ref{table erasure game value}.  If Player B does not know the state of the game, then on average A will get a payoff of 1/2.  If Player B does know the state $S$ then he will choose the strategy $B=S$, resulting in a payoff of zero.

\begin{table}
\caption{Value of erasure game under varying information structures}
\begin{center}
\label{table erasure game value}
\begin{tabular}{l|r}
& Value \\
\hline
\hline
Neither knows $S$ & 1/2 \\
Both know $S$ & 3/4 \\
A knows $S$ & 3/2 \\
B knows $S$ & 0
\end{tabular}
\end{center}
\end{table}

On the other hand, if Player A knows the state $S$ while Player B does not, then equilibrium will occur with Player A choosing $A=S$ and scoring $3/2$ on average.  Finally, we can consider the case where they both know the state.  As we observed in the game of Figure \ref{figure simple game}, Player A will choose the mixed strategy consisting of $A=S$ with probability 1/4 and $A=e$ with probability 3/4.  Player B will choose $B=S$ with probability 1/4 and $B=1-S$ with probability 3/4.  The resulting average payoff is 3/4.

In the ``erasure game'' the state is binary, so these four information structures mentioned exhaust all deterministic information structures.  No intriguing optimization problem presents itself in only one iteration of the game.  For example, we can't ask, ``what is the best information structure for Player A that has cardinality two?''  There is only one choice of information structure.  Fortunately, a more graceful spectrum of information structures is available with vector quantization.

\section{Shortcomings of Deterministic Encoding}

Consider a repeated-game setting of the erasure game where a helper observes the state of the game and communicates to Player A over a rate-limited channel.  Both players know all past actions and states.  Additionally, the helper, and possibly Player B, observe the state completely and non-causally.  The helper and Player A select a block length $n$ and arrange a protocol by which the helper will send $nR$ bits describing the state to Player A.  Player A will then play the game for $n$ iterations.  Player B has full knowledge of the protocol but does not actually see the message.  We ask for the maximum average payoff that can be achieved for a give rate $R$.  In other words, what is the max-min value of this game as a function of $R$?

We begin with a simple case.  As Table \ref{table erasure game value} shows, if Player A and Player B both know the state $S$, the value of the game is 3/4.  But what if Player A only learns of the state through communication from a helper, while Player B observes the state directly?  What rate of communication is needed for Player A to still achieve an average payoff of 3/4?  Recall that the uniquely optimal strategy for Player A is to choose $A=S$ with probability 1/4 and $A=e$ with probability 3/4.

We can use rate-distortion theory as inspiration for answering this question.  Rate-distortion theory prescribes a formula for finding the minimum description rate needed to reconstruct a sequence of observations with limited average distortion.  The procedure is to find the correlation of the source and reconstruction that satisfies the distortion constraint and results in the lowest mutual information.  By describing a random source at a rate greater than the mutual information, it is possible to assure with high probability that the reproduction will have the desired correlation with the source as measured by first order statistics.

In the case of the erasure game where Player B knows the state $S$, suppose we decide to encode the state for Player A using a rate-distortion-like code and an erasure test-channel such that the action sequence results in $S=A$ roughly 1/4 of the time and $S=e$ roughly 3/4 of the time.  The rate required is $R > I(S;A) = 1/4$ bits per iteration.  Unfortunately, this will not result in a good strategy for Player A.  The encoding schemes that arise in rate-distortion theory are deterministic.  The action sequence $A_1,...,A_n$ is a deterministic function of the state sequence $S_1,...,S_n$.  Since Player B observes the state sequence, he will deduce the actions of Player A and anticipate them every time.  Player A is effectively not playing a mixed strategy.  The resulting payoff will be 0.  The insufficiency of rate-distortion-like codes is a concern even when Player B does not know the state.  After watching the actions of Player A for roughly $k = \frac{R}{H(A)} n$ iterations it will be possible to deduce the entire action sequence.  To a clever opponent who does not know the state, the actions will appear random and appropriately distributed for the beginning $\frac{k}{n} < \frac{R}{H(A)} - \epsilon$ portion of the block, for large enough $n$ (related to results from \cite{Cuff2008}), and later in the block, when $\frac{k}{n} > \frac{R}{H(A)} + \epsilon$, the opponent will be able to decode the sequence with high probability and anticipate every action.  The bottom line is that rate-distortion-like codes place no emphasis on producing random actions.

\begin{figure}
\psfrag{l1}[][][0.8]{$S^n$}
\psfrag{l2}[][][0.8]{$U^n$}
\psfrag{l3}[][][0.8]{$A^n$}
\psfrag{s1}[][][.6]{x}
\psfrag{x}[][][.4]{x}
\psfrag{y}[][][.4]{x}
\psfrag{z}[][][.4]{x}
\psfrag{s1}[][][.4]{}
\centering
\includegraphics[width=0.35\textwidth]{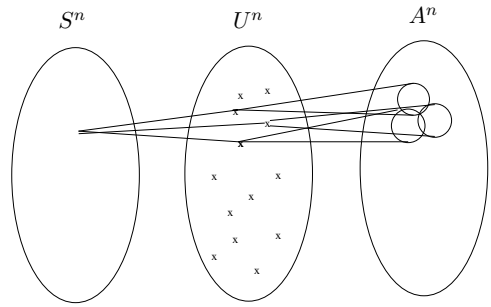}
\caption{{\em Encoding diagram.}  This illustrates a method of encoding a state sequence $S^n$ to allow an action sequence $A^n$ to be random even to an observer of the state.  An auxiliary variable $U$ is chosen which separates $S$ and $A$ into a Markov chain, and a codebook of $U^n$ sequences is agreed upon.  The state sequence $S^n$ is a random realization.  The encoder observes $S^n$ and randomly chooses among jointly typical $U^n$ sequences from the codebook.  After sending the index of the $U^n$ sequence to the decoder, an action sequence is generated randomly conditioned on $U^n$.  If the codebook is populated densely enough, the actions $A^n$ will be memoryless and appropriately correlated with the state sequence $S^n$.  The required density of the codebook is a topic visited in \cite{Cuff2008}, \cite{Wyner1975}, \cite{Cuff2009}, and \cite{Han}.}
\label{figure encoding diagram}
\end{figure}

The work in \cite{Cuff2008} prescribes an encoding scheme for generating correlated random variables.  The minimum description rate for the state sequence $S_1,...,S_n$ needed to produce a sequence of actions $A_1,...,A_n$ with a distribution that is arbitrary close in total variation to the desired mixed strategy is Wyner's common information $C(S;A)$, defined in \cite{Wyner1975}.  The resulting encoding scheme for producing this ``strong coordination'' \cite{Cuff2009} between the state $S$ and the action $A$ uses randomized encoding and randomized decoding.  Figure \ref{figure encoding diagram} illustrates that the encoder uses the message to specify the index of a sequence $U_1,...,U_n$ from a predefined codebook, just as in rate-distortion-like codes, but here the sequences do not represent reconstruction sequences.  After the decoder identifies the sequence $U_1,...,U_n$ he produces the actions $A_1,...,A_n$ randomly as the output of a memoryless channel from $U$ to $A$.  In this way the codebook is separated from the action sequence $A_1,...,A_n$, allowing more randomness to be injected into the actions.

The common information for a binary erasure channel can be found in \cite{Cuff2008}.  Thus, the minimum description rate needed for a helper to describe the state $S$ to Player A in order to achieve the optimal average payoff in the erasure game when Player B knows the state is $R > C(S;A) = H(1/4)$, where $H(\cdot)$ is the binary entropy function.

\section{Stationary Mixed Strategy not Necessary}

Just as the information structure in a Bayesian game defines the set of pure strategies for each player, the description rate of the state that a helper is allowed to use also defines the set of block strategies allowed, which may have structure and memory from one game iteration to the next.  We know from implications in \cite{Cuff2008} that the set of achievable stationary strategies is the set of conditional distribution $p_{A|S}$ that yield $C(S;A) < R$.  However, stationary strategies are not the only strategies worth considering.

\begin{figure}
\psfrag{a}[][][0.8]{0}
\psfrag{b}[][][0.8]{-1}
\psfrag{c}[][][0.8]{0}
\psfrag{d}[][][0.8]{0}
\psfrag{e}[][][0.8]{1}
\psfrag{f}[][][0.8]{Player B}
\psfrag{g}[][][0.8]{Player A}
\psfrag{h}[][][0.8]{}
\psfrag{i}[][][0.8]{$S=0$}
\psfrag{j}[][][0.8]{$S=1$}
\centering
\includegraphics[width=.4\textwidth]{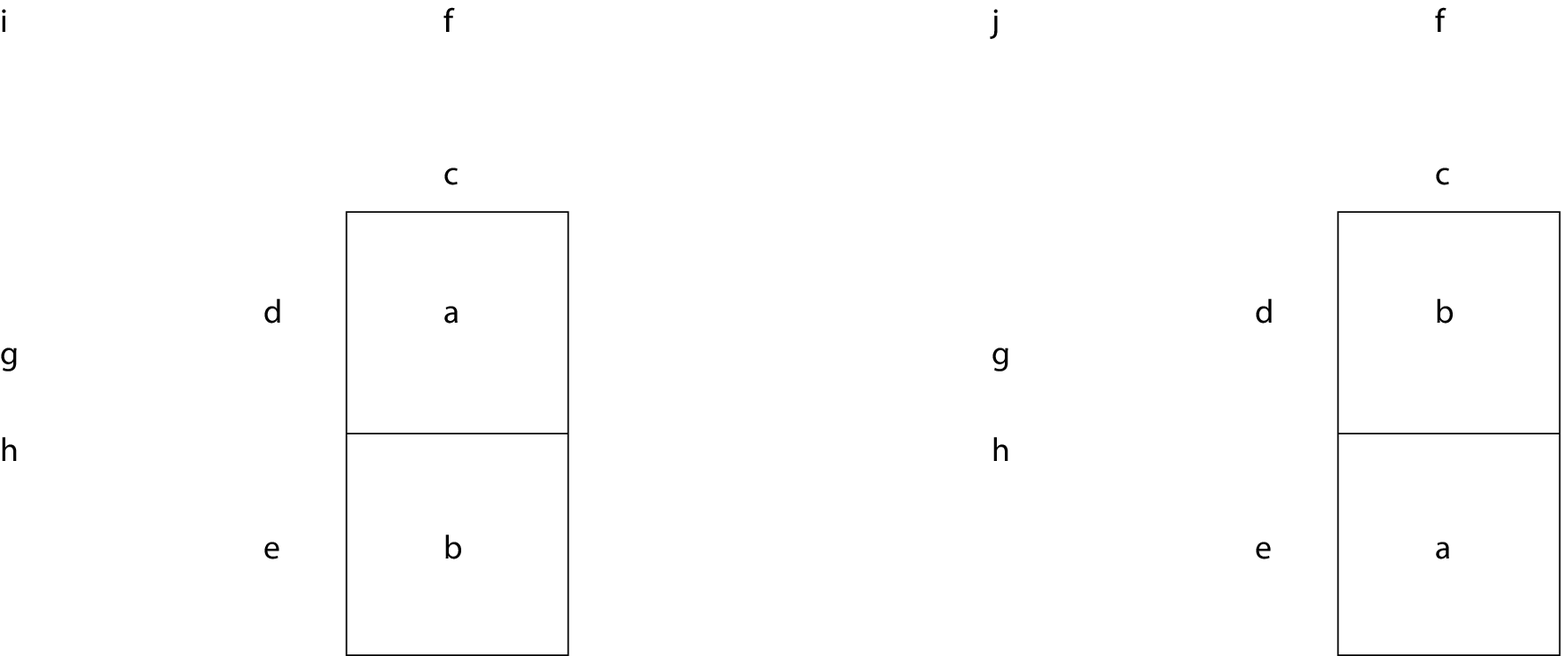}
\caption{{\em Degenerate game.}  This game has two states that are equally likely.  Player B does not have a choice of alternative strategies; therefore, from Player A's point of view this is not really an adversarial setting.  The payoff of the game depends only on the probability with which $A=S$.  The rate-value tradeoff of this degenerate game can be cast as a rate-distortion problem.  Randomized actions for Player A are unnecessary.}
\label{figure degenerate game}
\end{figure}

A degenerate Bayesian game is represented by the payoff matrices in Figure \ref{figure degenerate game}.  In this game Player B has only one pure strategy.  Player A is not really playing against an opponent but is simply maximizing the average value of a function of the state $S$ and action $A$.  In fact, the payoffs in this game are simply the negative Hamming distortion between $S$ and $A$; therefore, the rate-value tradeoff for this game is a canonical example from rate-distortion theory.  A payoff $P$ is achievable at rate $R$ if $R > 1 - H(-P)$ for $P \geq -1/2$, where $H(\cdot)$ is the binary entropy function.

In this degenerate Bayesian game the optimal strategy has structure and predictability.  There is no adversary to compete with, so producing a random sequence of actions is unnecessary.  These degenerate games simplify to rate-distortion problems and poignantly highlight situations where fully generating correlated random variables is not necessary.

\section{The Search for a Graceful Transition}

The state of a Bayesian game should be encoded by a helper in such a way that allows a random mixed strategy to be correlated with the state, even if that strategy is not entirely unpredictable for the duration of the communication block.  One way to achieve this is to follow the encoding procedure used to generate correlated random variables, as in Figure \ref{figure encoding diagram}, but use a smaller codebook.

Specifically, a codebook is constructed by first choosing a conditional distribution $p_{U,A|S} = p_{U|S}p_{A|U}$ and generating $2^{nR}$ sequences $u^n(i)$ independently for each $i$ and i.i.d. according to $p_U$.  The encoder chooses randomly (according to the appropriate distribution prescribed in \cite{Cuff2008}) from all sequences $u^n(i)$ that are jointly typical with the state and sends the index $I$ to the decoder.  The decoder then constructs the sequence $u^n(i)$ from the index $I$ and synthesizes a memoryless channel according to $p_{A|U}$ to produce an action sequence $A_1,...,A_n$.  If the rate $R > I(U;S,A)$ then this would produce a memoryless strategy, meaning that the opponent could not use observations of the past actions and states to infer about future actions.  However, by letting $R < I(U;S,A)$ we also allow for situations where memoryless strategies are not of the essence.

In order for the encoder to find at least one jointly typical $u^n(i)$ sequence in the codebook with high probability, a rate requirement of $R > I(S;U)$ is necessary.  Beyond that, any excess rate will serve to randomize the actions for a portion of the encoding block.  At first the actions will appear random and correlated with the state.  After observing a designated fraction of the block, the opponent will be able to deduce the index of the message using a channel decoder, revealing the future of the $u^n(i)$ sequence, from which the future actions will be randomly generated.  The transition from Player B knowing nothing about the next action $A$ to the point where Player B can decode $u^n(i)$ with high probability becomes sharper as the block length $n$ grows.

Let $\alpha$ be the transition threshold such that for the first $k$ iterations where $k < (\alpha - \epsilon)n$ the actions are random according to the mixed strategy $p_{A|S}$ and for the final $k$ iterations where $k > (\alpha + \epsilon)n$ the sequence $u^n(i)$ is known with high probability, in the limit as $n$ is large.  Then for the case where Player B does not know the state sequence, $\alpha = \frac{R}{I(U;S,A)}$.  For the case where Player B knows the state sequence, $\alpha = \frac{R - I(U;S)}{I(U;A|S)}$.

Let us designate some notation to represent the score Player A achieves in a game under different settings.  Let $\underline{\Pi}_{p_{A|S}}$ represent the minimum average payoff Player A receives by playing a strategy $p_{A|S}$ when Player B does not know the state $S$.  When Player B does know the state, we use a superscript to indicate this and represent the minimum average payoff for Player A by $\underline{\Pi}_{p_{A|S}}^{(S)}$.  Additionally, when an auxiliary random variable $U$ is involved in constructing the action $A$ such that $p_{U,A|S} = p_{U|S}p_{A|U}$ and Player B knows $U$, we represent the minimum average payoff for Player B as $\underline{\Pi}_{p_{A|U}}^{(U)}$ or $\underline{\Pi}_{p_{A|U}}^{(S,U)}$, depending on whether or not Player B also knows the state.  Each of these values can be computed from the payoff matrix $\Pi$:
\begin{eqnarray*}
\underline{\Pi}_{p_{A|S}} & \triangleq & \min_{p_{B}} \sum_{a,b,s} p_S(s)p_{A|S}(a|s)p_{B}(b) \Pi(a,b,s), \\
\underline{\Pi}_{p_{A|S}}^{(S)} & \triangleq & \min_{p_{B|S}} \sum_{a,b,s} p_S(s)p_{A|S}(a|s)p_{B|S}(b|s) \Pi(a,b,s), \\
\underline{\Pi}_{p_{A|U}}^{(U)} & \triangleq & \min_{p_{B|U}} \sum_{a,b,s,u} p_U(u)p_{S|U}(s|u)p_{A|U}(a|u) \ldots \\
& & \;\;\;\;\;\;\;\; p_{B|U}(b|u) \Pi(a,b,s), \\
\underline{\Pi}_{p_{A|U}}^{(S,U)} & \triangleq & \min_{p_{B|S,U}} \sum_{a,b,s,u} p_U(u)p_{S|U}(s|u)p_{A|U}(a|u) \ldots \\
& & \;\;\;\;\;\;\;\; p_{B|S,U}(b|s,u) \Pi(a,b,s).
\end{eqnarray*}

\begin{theorem}[Lower bound on rate-value tradeoff]
\label{theorem main}
If Player B does not know the state sequence, an average payoff $P$ for Player A is achievable with a state-information description rate of $R$ if
\begin{eqnarray*}
\exists (U,A) & \sim & p_{U,A|S} = p_{U|S}p_{A|U} \mbox{ such that} \\
\underline{\Pi}_{p_{A|S}} & > & P, \\
R & > & I(U;S), \\
R & > & I(U;S,A) \frac{P - \underline{\Pi}_{p_{A|U}}^{(U)}}{\underline{\Pi}_{p_{A|S}} - \underline{\Pi}_{p_{A|U}}^{(U)}}.
\end{eqnarray*}
If Player B knows the state sequence, an average payoff $P$ for Player A is achievable with a state-information description rate of $R$ if
\begin{eqnarray*}
\exists (U,A) & \sim & p_{U,A|S} = p_{U|S}p_{A|U} \mbox{ such that} \\
\underline{\Pi}_{p_{A|S}}^{(S)} & > & P, \\
R & > & I(U;S) + I(U;A|S) \left( \frac{P - \underline{\Pi}_{p_{A|U}}^{(S,U)}}{\underline{\Pi}_{p_{A|S}}^{(S)} - \underline{\Pi}_{p_{A|U}}^{(S,U)}} \right)^+.
\end{eqnarray*}
\end{theorem}

The lower bound of Theorem \ref{theorem main} accommodates settings where full randomization is needed, such as optimal play in the erasure game when Player B knows the state, and it also accommodates efficient communication for degenerate games where the payoff does not change when the opponent learns about the action $A$.  It is not clear, however, whether or not this gives the whole tradeoff for sub-optimal play in games where the communication limits do not allow for ideal randomization.  For example, what is the value of the erasure game when Player B knows the state sequence and $R < H(1/4)$, where $H(\cdot)$ is the binary entropy function?

If the bound in Theorem \ref{theorem main} is not tight, a couple of ideas come to mind for improvement.  One is to use an encoding method that is not stationary but moves from one strategy to another as the opponent learns.  In game settings, time-sharing must be analyzed carefully.  The performance depends on whether the time-sharing is interleaved or not.  A related idea is to use a layered encoding scheme, so that the opponent learns the message a little bit at a time.  A possible layered approach follows.

The helper who observes the state can send a message in layers to Player A by selecting two auxiliary random variables $U_1$ and $U_2$ and an action $A$ such that $S - (U_1,U_2) - A$ form a Markov chain.  A codebook of $u_1^n$ sequences of size $2^{n (I(U_1;S) + \epsilon)}$ is generated from the i.i.d. distribution, and for each of these sequences a codebook of $u_2^n$ sequences of size $2^{n (R - I(U_1;S) - \epsilon)}$ is generated conditioned on $u_1^n$.  We let $\epsilon$ be small.  The encoder first finds a $u_1^n$ sequence that is appropriately correlated with the state sequence $S^n$ and then chooses randomly from the $u_2^n$ sequences in the conditional codebook that are appropriately correlated with both $u_1^n$ and $S^n$.  The decoder constructs $u_1^n$ and $u_2^n$ from the message and synthesizes a memoryless channel to generate the action sequence $A^n$ from $u_1^n$ and $u_2^n$.

Using this encoding scheme, the block will be divided into three sections partitioned by $\alpha_1$ and $\alpha_2$, where the knowledge of Player B transitions sharply at these thresholds in the limit as $n$ becomes large.  For the first $k$ iterations where $k < (\alpha_1 - \epsilon)n$ the actions are random according to the mixed strategy $p_{A|S}$.  For the middle $k$ iterations where $(\alpha_1 + \epsilon)n < k < (\alpha_2 - \epsilon)n$ the opponent knows $U_1$, which is correlated with $A$, and for the final $k$ iterations where $k > (\alpha_2 + \epsilon)n$ the opponent knows both $U_1$ and $U_2$.  In the setting where Player B does not know the state sequence, $\alpha_1 = \frac{I(U_1;S)}{I(U_1;S,A)}$ and $\alpha_2 = \frac{R - I(U_1;S)}{I(U_2;S,A|U_1)}$ as long as $\alpha_1 < \alpha_2 < 1$.  In the setting where Player B knows the state sequence, $\alpha_1 = 0$ and $\alpha_2 = \frac{R - I(U_1,U_2;S)}{I(U_2;A|U_1,S)}$ as long as $\alpha_2 < 1$.\footnote{In the case where Player B does not know the state, if $\alpha_1 > \alpha_2$ then this layered encoding scheme does not provide any benefit over Theorem \ref{theorem main}. And with either assumption about Player B, if $\alpha_2 > 1$, the encoding scheme can be modified to increase $\alpha_1$.}

We omit the explicit lower bound that is obtained from this encoding scheme.  It is a straightforward derivation and provides little additional insight.  However, if a layered scheme proves to provide improvement over Theorem \ref{theorem main}, then the idea of adding additional layers with additional auxiliary variables comes to mind.  Each additional variable would introduce a new phase of performance in the game.  This quickly becomes cumbersome.  Perhaps instead there is a smooth way of adjusting the strategy as the game proceeds and the opponent infers more about the communication.

\section{Summary}

We have considered partial state information in a Bayesian game from an optimization perspective.  If a limited amount of state information is passed by a helper to one of the players in a two player zero-sum repeated Bayesian game, how much can it increase the value of the game?  The description of the state can be used to correlate the actions in the game with the state.  In some settings, mutual information is the description rate needed to adequately correlate behavior.  But in the adversarial setting of games, a rate-distortion-like compression of the state information at a rate equal to the mutual information results in behavior that is predictable by the opponent.  On the other hand, Wyner's common information has been shown to be the description rate needed to fully correlate actions in a completely unpredictable way.  However, this can be more than necessary.

We introduce a communication scheme that performs efficiently (Theorem \ref{theorem main}) in the two extreme cases, where memoryless randomization is essential and where it is irrelevant.  Although the communication is based on i.i.d. codebooks, the performance in the game changes dramatically mid-way through the communication block as the opponent infers the compressed message.  The non-stationary performance of a stationary encoding scheme introduces new challenges in the quest for efficient compression.

\bibliographystyle{unsr}

\begin{thebibliography}{1}
\bibitem{vonNeumann1928}
J. Von Neumann. ``Zur Theorie der Gesellschaftsspiele.'' \emph{Mathematische Annalen}, 100 (1928): 295-320.
\bibitem{Gossner2006}
O. Gossner. ``Ability and Knowledge.'' to appear in \emph{Games and Economic Behavior}.
\bibitem{Gilboa1991}
I. Gilboa and E. Lehrer. ``The Value of Information - An Axiomatic Approach.'' \emph{Journal of Mathematical Economics}, 20 (1991): 443-459.
\bibitem{Shmaya2006}
E. Shmaya. ``The Value of Information Structures in Zero-sum Games with Lack of Information on One Side.'' \emph{Internation Journal of Game Theory}, 34 (2006): 155-165.
\bibitem{Lehrer2006}
E. Lehrer and D. Rosenberg. ``What restrictions do Bayesian games impose on the value of information?'' \emph{Journal of Mathematical Economics}, 42 (2006): 343-357.
\bibitem{Cuff2008}
P. Cuff. ``Communication Requirements for Generating Correlated Random Variables.'' ISIT 2008, Toronto.
\bibitem{Wyner1975}
A. Wyner. ``The Common Information of Two Dependent Random Variables.'' \emph{IEEE Trans. on Info. Theory}, vol. IT-21, no. 2, March 1975.
\bibitem{Cuff2009}
P. Cuff, H. Permuter, and T. Cover. ``Coordination Capacity.'' submitted to \emph{IEEE Trans. on Info. Theory}, 2009.
\bibitem{Han}
T. Han and S. Verd\'{u}, ``Approximation Theory of Output Statistics,'' \emph{IEEE Trans. on Info. Theory}, vol. 39, no. 3, May 1993.
\end{thebibliography}

\end{document}